\documentclass{aa}
\usepackage{graphicx}
\usepackage{natbib}
\usepackage{scalerel}
\usepackage{multirow}   
\usepackage{mathrsfs}
\usepackage[table]{xcolor}
\usepackage{fontawesome}

\newcommand{\concept}{{\textsc{C}\textsc{o\scalerel*{\textbf{\textit{N}}}{n}cept}}}

\newcommand{\msun}{M_{\odot}\,h^{-1}}

\newcommand{\de}{\text{d}}

\usepackage{txfonts}
\usepackage[pdfencoding=auto,psdextra]{hyperref}
\hypersetup{
    colorlinks=true,
    linkcolor=blue,
    filecolor=magenta,      
    urlcolor=blue,
    citecolor=blue
}
\usepackage[utf8]{inputenc}
\usepackage{orcidlink}
\begin{document}

   \title{DUCA: Dynamic Universe Cosmological Analysis.}
   \titlerunning{The halo mass function in dynamical dark energy cosmologies}
   \authorrunning{Castro et al.}

   \subtitle{I. The halo mass function in dynamical dark energy cosmologies}

   \author{Tiago~Castro\orcidlink{0000-0002-6292-3228}
   \inst{1,2,3,4}
   \thanks{\email{tiago.batalha@inaf.it}}, 
   Stefano~Borgani\orcidlink{0000-0001-6151-6439}
   \inst{5, 1, 2, 3, 4}, and
   Jeppe~Dakin \orcidlink{0000-0002-2915-0315}
   \inst{6}
   }

   \institute{
   $^{1}$ INAF -- Osservatorio Astronomico di Trieste, Via G. B. Tiepolo 11, 34143 Trieste, Italy\\
    $^{2}$ INFN --  Sezione di Trieste, Via Valerio 2, 34127 Trieste TS, Italy\\
    $^{3}$ IFPU -- Institute for Fundamental Physics of the Universe, via Beirut 2, 34151 Trieste, Italy\\
    $^{4}$ ICSC -- Via Magnanelli 2, Bologna, Italy\\
    $^{5}$ Dipartimento di Fisica -- Sezione di Astronomia, Università di Trieste, Via Tiepolo 11, Trieste, 34131, Italy\\
    $^{6}$ Department of Astrophysics -- University of Zurich, Winterthurerstrasse 190, 8057 Z{\"u}rich, Switzerland
             }


 
  \abstract 
  {The halo mass function (HMF) is fundamental for interpreting the number counts of galaxy clusters, serving as a pivotal theoretical tool in cosmology. With the advent of high-precision surveys such as LSST, eROSITA, DESI, and Euclid, accurate HMF modeling becomes indispensable to avoid systematic biases in cosmological parameter estimation from cluster cosmology. Moreover, these surveys aim to shed light on the dark sector and uncover dark energy's puzzling nature, necessitating models that faithfully capture its features to ensure robust parameter inference.} 
  {We aim to construct a model for the HMF in dynamical dark energy cosmologies that preserves the accuracy achieved for the standard $\Lambda (\nu)$CDM model of cosmology, while meeting the precision requirements necessary for future cosmological surveys.} 
  {Our approach models the HMF parameters as functions of the deceleration parameter at the turnaround, a quantity shown to encapsulate essential information regarding the impact of dynamical dark energy on structure formation. We calibrate the model using results from a comprehensive suite of $N$-body simulations spanning various cosmological scenarios, ensuring sub-percent systematic accuracy.} 
  {We present an HMF model tailored for dynamical dark energy cosmologies. The model is calibrated following a Bayesian approach, and its uncertainty is characterized by a single parameter controlling its systematic error, which remains at the sub-percent level. This ensures that theoretical uncertainties from our model are subdominant relative to other error sources in future cluster number counts analyses.}
   {}


   \maketitle
%

\section{\label{sec:intro}Introduction}

Structure formation in the Universe proceeds hierarchically, with small-scale perturbations collapsing and merging to form larger structures over cosmic time. Galaxy clusters, being the most massive virialized objects, sit at the apex of this hierarchy and serve as powerful probes of cosmology~\citep[see reviews by][]{Allen:2011zs, KravtsovBorgani:2012}. The abundance and distribution of galaxy clusters provide valuable cosmological information, including the growth of cosmic structures and the nature of dark energy~\citep{Borgani:2001,Schuecker:2002ti,Majumdar:2003mw,Vikhlinin:2008ym,Planck:2015lwi,Marulli:2018owk,SPT:2018njh,DES:2020cbm,Fumagalli:2023yym}.

The halo mass function (HMF), which describes the comoving number density of dark matter halos as a function of mass and redshift, is a key theoretical tool for interpreting observations of galaxy clusters~\citep[e.g.,][]{Press:1973iz, Bond:1990iw, Sheth:1999mn, Sheth:1999su, Tinker:2008ff, Despali:2015yla, Bocquet:2020tes, Ondaro-Mallea:2021yfv, Euclid:2022dbc}. Accurate modeling of the HMF is essential for not biasing the cosmological parameters constraints from cluster surveys~\citep[e.g.,][]{Salvati:2020exw, Artis:2021tjj}. 

Due to their non-linearity, the accurate and precise modeling of the HMF  requires simulations that fully capture the non-linear evolution of cosmic structures. $N$-body simulations~\citep[see][for a review]{Angulo:2021kes} provides theoretical means to examine the non-linear regime at which galaxy clusters are entangled. They operate, however, under the assumption that the baryonic feedback is subdominant to gravity. Yet, despite being a minor component in our Universe, different studies showed that luminous matter significantly affects structure formation in the Universe~\citep{Cui:2014aga,Velliscig:2014,Bocquet:2015pva,Castro:2020yes, Euclid:2023jih}. At the scale of galaxy clusters, it is well-understood that baryonic feedback does not disrupt structures; instead, it redistributes the halo's composition, altering its mass compared to the same object simulated with a collisionless scheme. Given that hydrodynamical simulations are substantially more computationally demanding than purely gravitational $N$-body simulations, the commonly adopted approach is to characterize the HMF using the latter and then model the impact of baryonic physics on halo masses in post-processing~\citep[see, e.g.,][]{Schneider:2015wta, Arico:2020lhq, Euclid:2023jih}. The baryonic implications in the HMF will not be addressed in the rest of this paper.

In the standard cosmological model, $\Lambda$ cold dark matter (CDM), the accelerated expansion of the Universe is driven by a cosmological constant ($\Lambda$). However, the physical nature of dark energy remains one of the biggest challenges in modern cosmology. Dynamical dark energy models~\citep[see, for instance,][]{Peebles:2002gy, Copeland:2006wr}, where the dark energy equation of state evolves with time, offer alternatives to the cosmological constant and can leave distinctive signatures on the formation and evolution of cosmic structures~\citep{Frieman:2008sn, Weinberg:2013agg}.

Ongoing and upcoming missions such as the Vera C. Rubin Observatory's Legacy Survey of Space and Time~\citep[LSST,][]{LSSTScience:2009jmu},\footnote{\url{https://www.lsst.org}} the third generation of the South Pole Telescope~\citep[SPT-3G,][]{SPT-3G:2014dbx},\footnote{\url{https://astro.fnal.gov/science/cmbr/spt-3g/}} the Dark Energy Spectroscopic Instrument~\citep[DESI,][]{DESI:2016fyo},\footnote{\url{https://www.desi.lbl.gov}} the Nancy Grace Roman Space Telescope~\citep{Spergel:2015sza},\footnote{\url{https://roman.gsfc.nasa.gov}} the Square Kilometre Array~\citep[SKA,][]{Maartens:2015mra},\footnote{\url{https://www.skatelescope.org}} eROSITA~\citep{eROSITA:2020emt},\footnote{\url{https://www.mpe.mpg.de/eROSITA}} and Euclid~\citep{euclidoverview}\footnote{\url{https://www.euclid-ec.org}} will provide unprecedented observations of the large-scale structure of the Universe and will help to shed light on the dark sector nature. These surveys will push the statistical uncertainties to never-seen levels, and theoretical models must keep pace.

Understanding how dynamical dark energy affects the HMF is, therefore, crucial for extracting accurate cosmological information from current and future surveys of galaxy clusters. Previous studies have explored the impact of different dark energy features on the HMF~\citep[e.g.,][]{Courtin:2010gx, Cui:2012is, Bhattacharya:2010wy, Saez-Casares:2024bzg, Shen:2024cio}, often finding that variations in its dark nature can lead to significant differences in the predicted abundance of massive halos, especially at low redshifts when the dynamics of the Universe dynamics is driven by dark energy. While these works have modeled the HMF in different scenarios, the diversity of dynamical dark energy models and the precision requirements of upcoming surveys necessitate further work to keep the pace in accuracy and precision of the available predictions for the HMF in the $\Lambda$CDM. 

In this work, we extend our previous study on the HMF~\citep{Euclid:2022dbc} to dynamical dark energy models described by the Chevallier--Polarski--Linder (CPL) parametrization~\citep{Chevallier:2000qy, Linder:2002et}. The CPL framework provides a widely used phenomenological model that describes the dark energy equation of state $w(a)$ as a linear function of the scale factor $a$ with two parameters: $w_0$ is the present-day value of $w_{\rm DE}$, and $w_a$ characterizes its evolution. We aim to calibrate a new model for the HMF that accounts for the effects of a time-varying dark energy equation of state while retaining the accuracy needed for the analysis of upcoming cluster surveys, ensuring that theoretical uncertainties do not dominate the error budget.

To this end, we perform a series of precise $N$-body simulations spanning various cosmologies. We analyze the resulting halo catalogs to quantify the impact of dynamical dark energy on the HMF and to develop a fitting function that accurately reproduces the simulation results across the parameter space.

This paper is organized as follows: in Sect.~\ref{sec:theory}, we present the theoretical framework, focusing on the $w_0w_a$ model and the HMF. In Sect.~\ref{sec:methodology}, we describe the methodology used for calibrating the HMF model, including the simulation setup and the Bayesian approach for parameter estimation. In Sect.~\ref{sec:results}, we present our results, highlighting the accuracy and robustness of the proposed model compared to other existing models. Our conclusions are drawn in Sect.~\ref{sec:conclusions}. Finally, the \texttt{python} implementation of our model is publicly available at~\url{https://github.com/TiagoBsCastro/CCToolkit} and presented in Sect.~\ref{sec:data}.


\section{\label{sec:theory}Theory}

In this section, we present a short overview of the main concepts of the CPL model (Sect.~\ref{sec:w0wa_model}) and the HMF (Sect.~\ref{sec:hmf}).

\subsection{\label{sec:w0wa_model}The $w_0w_a$CDM model}

The CPL model~\citep{Chevallier:2000qy,Linder:2002et} is a phenomenological approach to describe the evolution of the dark energy equation of state parameter $w_{\rm DE}$, that determines the relation between pressure and density of the DE, $p=w_{\rm DE}\rho c^2$. It is expressed as a function of the scale factor $a$ as
\begin{equation}
w_{\rm DE}(a) = w_0 + w_a (1 - a)\,,
\label{eq:w_of_a}
\end{equation}
or equivalently in terms of redshift
\begin{equation}
w_{\rm DE}(z) = w_0 + w_a \left( \frac{z}{1 + z} \right)\,,
\label{eq:w_of_z}
\end{equation}
where $w_0$ is the present-day value of the dark energy equation of state and $w_a$ its rate of change with respect to the scale factor.

In a flat universe including matter, radiation, massive neutrinos, and dark energy components, the Hubble parameter $H(z)$ is given by
\begin{align}
\frac{H^2(z)}{H_0^2} = \Omega_{\rm m,0} (1 + z)^3 + \Omega_{\rm r,0} (1 + z)^4+ \Omega_\nu (z)\,\frac{\rho_{\rm c} (z)}{\rho_{\rm c, 0}} \nonumber \\
+\Omega_{\rm DE,0} \, e^{3 \int_0^z \frac{1 + w(z')}{1 + z'} dz'}\,,
\label{eq:hubble_parameter}
\end{align}
where $H$ is the Hubble constant and $H_0$ its value at present-day; $\Omega_{\rm m,0}$, $\Omega_{\rm r,0}$, and $\Omega_{\rm DE,0}$ are the present-day density parameter for matter (including both baryonic and cold dark matter), radiation, and dark energy. $\Omega_\nu(z)$ is the neutrino density parameter at redshift $z$, $\rho_{\rm c}(z)$ is the critical density at redshift $z$ and $\rho_{\rm c, 0}$ its value at present-day. 

The integral in the exponential in Eq.~\eqref{eq:hubble_parameter} accounts for the evolution of dark energy density due to its dynamic equation of state. Using the $w_0w_a$ parametrization, the exponential term becomes \citep{Linder:2002et}
\begin{equation}
e^{3 \int_0^z \frac{1 + w(z')}{1 + z'} dz'} = (1 + z)^{3(1 + w_0 + w_a)} e^{-3 w_a z / (1 + z)}\,.
\label{eq:dark_energy_evolution}
\end{equation}

Standard Massive neutrinos play a unique role in cosmology, affecting both the background expansion and the growth of cosmic structures \citep{Lesgourgues:2006nd}. The present-day value of the neutrino density parameter $\Omega_{\nu, 0}$ depends on the sum of the masses of the three neutrino species $\sum m_\nu$ and is approximately given by
\begin{equation}
\Omega_{\nu,0} h^2 = \frac{\sum m_\nu}{93.14\, \text{eV}}\,,
\label{eq:omega_nu}
\end{equation}
where $h = H_0 / (100\, \rm{km}\, \rm{s}^{-1}\, \rm{Mpc}^{-1})$. At early times ($z \gg 1$), neutrinos behave like radiation ($w_\nu = \tfrac{1}{3}$), while at late times ( $z \lesssim 1 $), they become non-relativistic and act like a pressureless fluid ($w_\nu = 0 $). Lastly, massive neutrinos are known to follow a hierarchy in their masses, with possible normal or inverted ordering~\citep{Lesgourgues:2006nd}; however, for simplicity, we assume in the rest of this work that all three neutrino species have the same mass.

The main feature of dark energy models is explaining the current observed accelerated expansion of the Universe. The deceleration parameter $q_{\rm dec}$ the variation of the expansion rate and is defined as \citep{Weinberg:2008zzc, peebles2020large}
\begin{equation}
q_{\rm dec} = -\frac{\ddot{a}}{a H^2}\,,
\label{eq:deceleration_parameter}
\end{equation}
where the dots denote derivatives with respect to cosmic time.

Using the Friedmann equations, $q_{\rm dec}$ can be expressed in terms of the density parameters and their respective equations of state:
\begin{equation}
q_{\rm dec}(z) = \frac{1}{2} \sum_i \Omega_i(z) \left[ 1 + 3 w_i(z) \right]\,,
\label{eq:q_with_omega}
\end{equation}
where the sum runs over all components $i$ (matter, radiation, neutrinos, dark energy), and $w_i(z)$ is the equation of state parameter for each component:
\begin{itemize}
    \item $w_m = 0$ for matter.
    \item $w_r = \tfrac{1}{3}$ for radiation.
    \item $w_\nu(z)$ for neutrinos, transitioning from $\tfrac{1}{3}$ to $0$.
    \item $w_{\rm{DE}}(z)$ given by the CPL model for dark energy.
\end{itemize}
At the present time (\( z = 0 \)), the deceleration parameter is approximately
\begin{equation}
q_{\rm dec,0} \approx \frac{1}{2} \left[ 2\,\Omega_{\rm m,0}  + \Omega_{\rm{DE},0} (1 + 3 w_0) \right]\,.
\label{eq:q0}
\end{equation}
Thus, a negative value of $q_{\rm dec}$ indicates cosmic acceleration, while a positive value indicates deceleration. Since the deceleration parameter is measured to be negative at the present time, a necessary condition for this acceleration is that the equation of state parameter satisfies $w_0<-1/3$.

\subsection{The Halo Mass Function}
\label{sec:hmf}

The comoving number density of dark matter halos within the mass range $[M, M+\de M]$ is described by the differential HMF
\begin{equation}
\frac{\de n}{\de M} \de M = \frac{\rho_{\rm m}}{M} \, \nu f(\nu) \, \de \ln \nu\,,
\label{eq:hmf}
\end{equation}
where $\rho_{\rm m}$ is the mean comoving matter density of the Universe, $\nu$ is the peak height parameter, and $\nu f(\nu)$ is known as the multiplicity function. The peak height $\nu$ quantifies the rarity of a halo and is defined as
\begin{equation}
\nu = \frac{\delta_{\rm c}}{\sigma(M, z)}\,,
\label{eq:nu}
\end{equation}
with $\delta_{\rm c}$ being the critical density threshold for spherical collapse linearly extrapolated to redshift $z$, and $\sigma(M, z)^2$ representing the mass variance at redshift $z$. The mass variance is computed from the linear matter power spectrum $P_{\rm m}(k, z)$ via:
\begin{equation}
\sigma^2(M, z) = \frac{1}{2\pi^2} \int_0^\infty \de k\,k^2\,P_{\rm m}(k, z)\,W^2(k R)\,,
\label{eq:sigma}
\end{equation}
where $R(M)$ is the Lagrangian radius corresponding to mass $M$, given by $R = \left( 3 M / 4\pi \rho_{\rm m} \right)^{1/3}$, and $W(k R)$ is the Fourier transform of the real-space top-hat window function.

In cosmologies with massive neutrinos, the structure formation is tightly connected to the cold dark matter plus baryons density field instead of the total density field, including the neutrino contribution~\citep{Castorina:2014,Costanzi:2013bha,Vagnozzi:2018pwo}. Therefore, in this scenario, it is necessary that the matter field considered in the comoving matter density field $\rho_{\rm m}$ in Eq.~\eqref{eq:hmf} and the matter power-spectrum $P_{\rm m}$ in Eq.~\eqref{eq:sigma} refers to the cold dark matter plus baryons only to evaluate the model correctly~\citep[see also][]{Euclid:2022dbc,Euclid:2024wog}.

The assumption of universality implies that the multiplicity function $\nu f(\nu)$ is independent of cosmology when expressed in terms of the peak height $\nu$. While this holds approximately, several studies using $N$-body simulations have identified systematic deviations from universality, especially at late times \citep{Crocce:2009mg, Courtin:2010gx, Watson:2012mt, Diemer:2020rgd, Ondaro-Mallea:2021yfv}.

These deviations have been attributed to factors such as the specific definition of halos \citep{Watson:2012mt, Despali:2015yla, Diemer:2020rgd, Ondaro-Mallea:2021yfv} and a residual cosmology dependence of the collapse threshold $\delta_{\rm c}$. For instance, \citet{Courtin:2010gx} demonstrated that accounting for the cosmological variation of $\delta_{\rm c}$ leads to a more universal HMF, particularly at the high-mass end, which is critical for cluster cosmology.

In our analysis, we define halos using the spherical overdensity (SO) criterion, where halos are spheres within which the mean density is $\Delta_{\rm vir}(z)$ times the background matter density. The value of the virial overdensity $\Delta_{\rm vir}(z)$ and its dependence on cosmological parameters are determined by the spherical collapse model solution~\citep{Eke:1996ds}. 

We adopt the fitting formula for $\Delta_{\rm vir}(z)$ presented by~\citet{Bryan_1998}, even though it was calibrated for $\Lambda$CDM cosmologies and the results for dynamical dark energy models can differ significantly when derived from the numerical solution of the spherical collapse. The rationale behind this choice is threefold. First, this definition has already been implemented in several halo finders for practicality. Second, numerically solving the spherical collapse for arbitrary dynamical dark energy fluids poses several technical~\citep{Pace:2017qxv} and physical~\citep{Batista:2021uhb} challenges. Third, we aim for our HMF model to be computationally efficient so that it does not add computational load during cluster cosmology likelihood computations.  Nonetheless, we are confident that using the solution for the virial threshold would likely preserve the universality of the HMF more effectively than the adopted definition, as it has been observed previously for other definitions in $\Lambda$CDM cosmologies~\citep{Despali:2015yla, Diemer:2020rgd, Ondaro-Mallea:2021yfv}.

We adopt the fitting function for the multiplicity function introduced by \citet{Euclid:2022dbc} as a starting point for our modeling. This multiplicity function is given by
\begin{equation}
\nu f(\nu) = A(p, q) \sqrt{\frac{2 a \nu^2}{\pi}} \exp\left( -\frac{a \nu^2}{2} \right) \left[ 1 + \frac{1}{(a \nu^2)^p} \right] (\sqrt{a} \nu)^{q - 1}\,,
\label{eq:mult}
\end{equation}
where the parameters $a$, $p$, and $q$ are functions of the cosmological background evolution and the shape of the matter power spectrum. Specifically, they are modeled as:
\begin{align}
a &= a_R \, \Omega_{\rm m}^{a_z}(z)\,, \label{eq:a_param} \\
p &= p_1 + p_2 \left( \frac{\de \ln \sigma}{\de \ln R} + 0.5 \right)\,, \label{eq:p_param} \\
q &= q_R \, \Omega_{\rm m}^{q_z}(z)\,, \label{eq:q_param}
\end{align}
with
\begin{align}
a_R &= a_1 + a_2 \left( \frac{\de \ln \sigma}{\de \ln R} + 0.6125 \right)^2\,, \label{eq:aR_param} \\
q_R &= q_1 + q_2 \left( \frac{\de \ln \sigma}{\de \ln R} + 0.5 \right)\,. \label{eq:qR_param}
\end{align}
Here, $\de \ln \sigma / \de \ln R$ is the mass variance's logarithmic slope as a function of the Lagrangian radius $R$.

The normalization constant $A(p, q)$ is determined by the requirement that all matter is contained within halos, as prescribed by the halo model \citep{Cooray:2002dia}. When the parameters $a$, $p$, and $q$ are independent of $\nu$, the normalization is given by
\begin{equation}
A(p, q) = \left\{ \frac{2^{-1/2 - p + q/2}}{\sqrt{\pi}} \left[ 2^p \Gamma\left( \frac{q}{2} \right) + \Gamma\left( -p + \frac{q}{2} \right) \right] \right\}^{-1}\,,
\label{eq:Anorm}
\end{equation}
where $\Gamma$ denotes the Gamma function. It should be noted, however, that this normalization condition may not strictly hold in the real Universe~\citep{Angulo:2009hf}. Our model uses Eq.~\eqref{eq:Anorm} to ensure appropriate asymptotic behavior when extrapolating the HMF to lower masses, even though the parameters vary with cosmology and scale.

The values for the parameters $\{a_1, a_2, a_z, p_1, p_2, q_1, q_2, q_z\}$ are taken from Table 4 of \citet{Euclid:2022dbc}, calibrated using the \textsc{Rockstar} halo finder. We refer to that paper for a thorough discussion on the effect of changing the halo finder on the resulting HMF.

Finally, for the cosmology-dependence of the critical density threshold $\delta_{\rm c}$, we employ the fitting formula provided by \citet{Kitayama:1996ne}. 


\section{\label{sec:methodology}Methodology}

In this Section, we present the methodology used to calibrate the HMF model, including the presentation of the simulations used (Sect.~\ref{sec:sims}), the halo finder (Sect.~\ref{sec:halofinder}), and the Bayesian analysis (Sect.~\ref{sec:cal}).

\subsection{\label{sec:sims}Simulations}

\begin{figure*}
    \centering
    \includegraphics[width=1.0\textwidth]{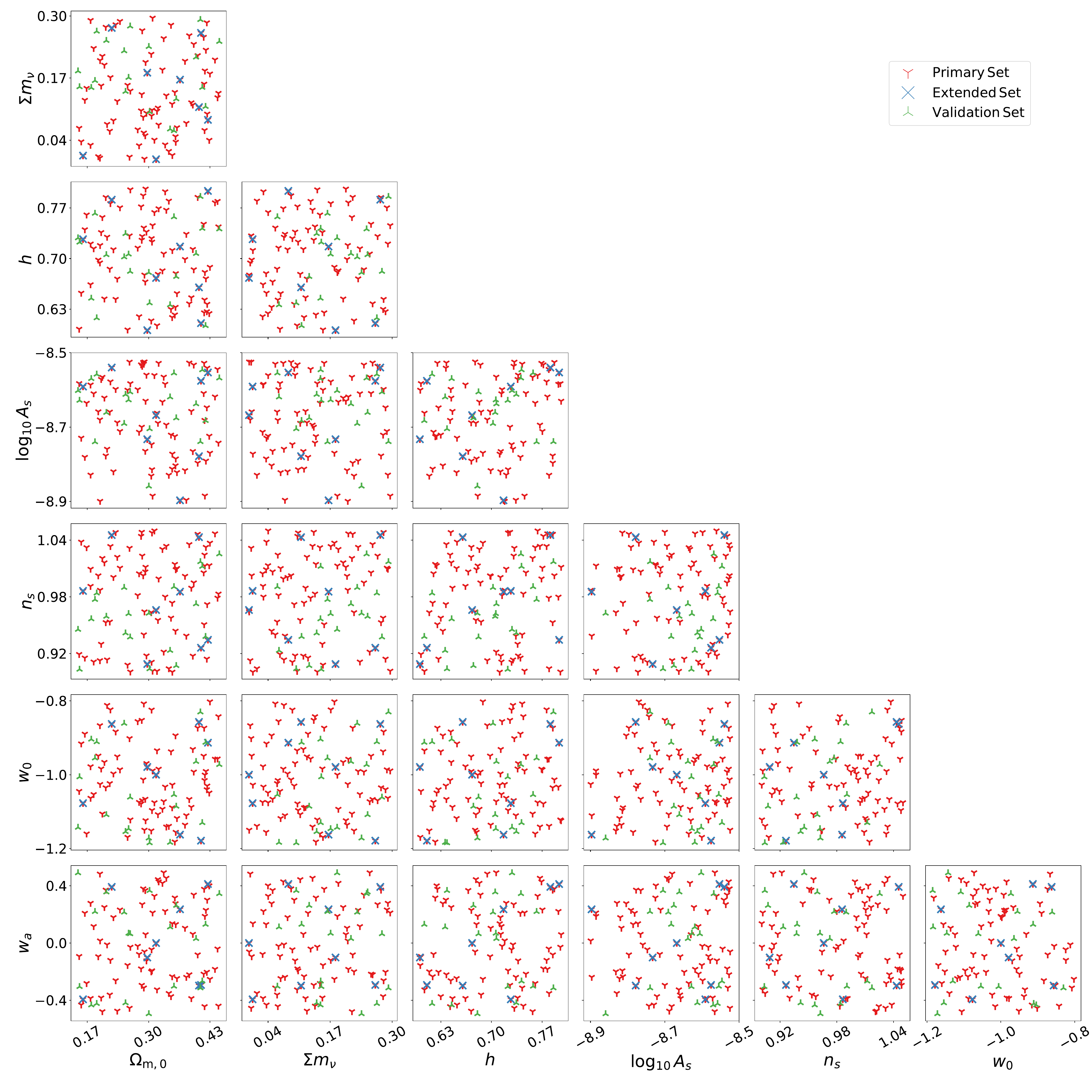}
    \caption{Cosmological parameters for the three sub-sets of the DUCA simulations. See Sect.~\ref{sec:sims} for details.}
    \label{fig:cosm}
\end{figure*}

This work utilizes the \textit{Dynamic Universe Cosmological Analysis} (DUCA) cosmological $N$-body simulation suite. This set of simulations was generated using the \concept{} code~\citep{Dakin:2021ivb},\footnote{\url{https://github.com/jmd-dk/concept/}} with detailed documentation available online.\footnote{\url{https://jmd-dk.github.io/concept/}} 

Initial conditions have been created on the fly by \concept{} using the recently implemented third-order Lagrangian perturbation theory (3LPT) module at the initial redshift $z_{\rm in}=24$. We adjusted the force split range from their default values in \concept{} to achieve higher accuracy in our simulations. Table~\ref{tab:simulation_params} lists the specific force parameters used in our simulations~\citep[see][for the precise definitions of each parameter]{Dakin:2021ivb}.
\begin{table*}[h!]
\centering
\caption{Force parameters used in the DUCA set of simulations, including short-range and long-range force specifications.}
\label{tab:simulation_params}
\begin{tabular}{lll}
\hline
\hline
\textbf{Parameter} & \textbf{Value} & \textbf{Description} \\
\hline
\multicolumn{3}{l}{\textbf{Short-Range Force Parameters}} \\
\texttt{scale}     & $1.25 \times \texttt{boxsize/gridsize}$ & Long/short-range force split scale \\
\texttt{range}     & $5.5 \times \texttt{scale}$             & Maximum reach of short-range force \\
\texttt{tablesize} & $2^{12}$                                & Size of tabulation for short-range forces \\
\hline
\multicolumn{3}{l}{\textbf{Long-Range Force Parameters}} \\
\texttt{gridsize\_pm} & $\sqrt[3]{\texttt{N}}$ & Particle-mesh (PM) grid size for gravity \\
\texttt{gridsize\_p3m} & $2 \times \sqrt[3]{\texttt{N}}$ & Particle-Particle Particle-Mesh (P3M) grid size for gravity \\
\hline
\end{tabular}
\tablefoot{The adopted setup for the DUCA simulation set splits the long-range forces into two grids. The particle-mesh (with grid size specified by \texttt{gridsize\_pm}) is used to continually realize the massive neutrinos and dark energy fluctuations according to GR. The particle-particle-particle-mesh (with grid size specified by \texttt{gridsize\_p3m}) is used for the gravity interaction between the CDM plus baryons component. We adopt grid sizes proportional to the cubic root of the total number of particles in the simulation \texttt{N}.}
\end{table*}

The DUCA simulation set was designed to explore the impact of varying cosmological parameters on the large-scale structure with high precision and efficiency and is divided into the three subsets we describe below and overview in Table~\ref{tab:simulation_sets}. We present the triangle plot with the values of the cosmological parameters assumed for $\Omega_{\rm m,0}$, $\sum m_\nu$, $h$, $\log_{10} A_{\rm s}$, $n_{\rm s}$, $w_0$, and $w_a$ by the different sub-sets in Fig.~\ref{fig:cosm}. The ranges within which these parameters vary were deliberately chosen to be conservative, encompassing the constraints from multiple cosmological probes. All simulations incorporate the effects of massive neutrinos and dark energy perturbations at background and linear perturbation levels through continual grid realization, with general relativistic effects fully accounted for~\citep{Dakin:2019vnj}. \concept{} also deploys a non-linear solver for the continuity and Euler equations for massive neutrinos on a grid following the approach outlined by~\citet{Dakin:2017idt}. However, operating under this setup results in significant computational overhead. Furthermore,~\citet{Euclid:2022qde} showed that the linear approach is sufficient to reproduce the impact of massive neutrinos in the HMF at the sub-percent level.

\begin{table*}[h!]
\centering
\caption{Summary of the simulation sets used in this work.}
\label{tab:simulation_sets}
\begin{tabular}{lcccccc}
\hline
\hline
\textbf{Simulation Set} & \textbf{Number} & \textbf{Box Size}       & \textbf{Particle Number} & \textbf{Mass Resolution} & \textbf{Initial Phases} & \textbf{Purpose} \\
                        & \textbf{of Sims} & $[h^{-1}\mathrm{Gpc}]$  &                          & $[(h/\Omega_{\rm m, 0})^{-1} M_\odot]$       &                        &                 \\
\hline
Primary Set             & 82 pairs        & $1$                      & $1024^3$                 & $\sim 2.6\times 10^{11}$            & Fixed and Paired       & Calibration     \\
Large Volume            & 8 pairs         & $2$                      & $2048^3$                 & $\sim2.6\times10^{11}$     & Fixed and Paired       & Calibration     \\
High Resolution         & 8 pairs         & $1$                      & $2048^3$                 & $\sim3.2\times10^{10}$   & Fixed and Paired       & Calibration     \\
Validation Set          & 20              & $1$                      & $1024^3$                 & $\sim2.6\times10^{11}$            & Random                & Validation      \\
\hline
\end{tabular}
\end{table*}

\subsubsection{Primary Simulation Set}

Our primary simulation set consists of 82 pairs of simulations, each corresponding to different cosmological parameters. The cosmological parameters that we varied are: $\Omega_{\rm m, 0}$, $\Omega_{\rm b, 0}$, $\sum m_\nu$, $H_0$, $n_{\rm s}$, $A_{\rm s}$, $w_0$ and $w_a$.

These parameters are sampled using Latin hypercube sampling to cover the multidimensional parameter space efficiently. Each simulation has a comoving box size of $1\, h^{-1}\, \mathrm{Gpc}$ and contains $1024^3$ dark matter particles, resulting in a particle mass resolution of approximately $2.6 \times 10^{11}\, (h/\Omega_{\rm m, 0})^{-1}\, M_\odot$. The simulations are evolved from redshift $z = 24$ to $z = 0$.

We employ fixed amplitudes and paired phases  in the Fourier representation of the density fluctuation field for initial condition generation, following the method described by~\citet{Angulo:2016hjd}. This technique generates pairs of simulations with initial power spectra which have identical amplitude, fixed at the mean $P(k)$ value, but with phases of the corresponding Fourier modes shifted by $\pi$. This approach effectively cancels the leading-order cosmic variance of clustering statistics when averaging over the pair.

\subsubsection{Extended Simulation Sets}

From the initial 82 cosmologies, we select eight cosmologies presented in Table~\ref{tab:simulation_cosmo}. They are indexed according to their value of $S_8=\sigma_8\sqrt{(\Omega_{\rm m, 0}/0.3)}$. The cosmological parameters were selected by brute-force searching for the cosmological set that presented the largest variance. For these chosen cosmologies, we perform additional simulations to refine our HMF modeling further:
\begin{itemize}
    \item Large-Volume Simulations: We run eight pairs of simulations with the same mass resolution but with a larger comoving box size of $2\, h^{-1}\, \mathrm{Gpc}$. This increases the simulation volume and number of particles by a factor of 8, thus enhancing the statistics for massive halos and reducing sample variance at the high-mass end. 

    \item High-Resolution Simulations: We run eight pairs of simulations with the same comoving box size and phases as the primary set ($1\, h^{-1}\, \mathrm{Gpc}$) but with an increased particle count of $2048^3$. This improves the mass resolution by a factor of 8, allowing us to probe smaller-mass halos and better resolve the low-mass end of the HMF. 
\end{itemize}

\begin{table*}[h!]
\centering
\caption{Cosmological parameters for the extended simulation set.}
\label{tab:simulation_cosmo}
\begin{tabular}{cccccccccc}
\hline\hline
\textbf{Index} & $\boldsymbol{\Omega_{\rm m,0}}$ & $\boldsymbol{\Omega_{\rm b,0}}$ & $\boldsymbol{\sum m_\nu}$ \textbf{[eV]} & $\boldsymbol{H_0}$ \textbf{[km/s/Mpc]} & $\boldsymbol{n_{\rm s}}$ & $\boldsymbol{A_{\rm s}}$ & $\boldsymbol{w_0}$ & $\boldsymbol{w_a}$ & $\boldsymbol{S_8}$ \\
\hline
0 & 0.2972 & 0.0974 & 0.1812 & 60.10 & 0.9089 & $1.8505 \times 10^{-9}$ & $-0.9791$ & $-0.1010$ & $0.4378$\\
1 & 0.1610 & 0.0302 & 0.0076 & 72.67 & 0.9863 & $2.5652 \times 10^{-9}$ & $-1.0770$ & $-0.3931$ & $0.5269$\\
2 & 0.2218 & 0.0438 & 0.2750 & 78.14 & 1.0454 & $2.8856 \times 10^{-9}$ & $-0.8626$ & $0.3918$ & $0.6340$\\
3 & 0.3158 & 0.0494 & 0.0000 & 67.32 & 0.9661 & $2.1488 \times 10^{-9}$ & $-1.0000$ & $0.0000$ & $0.8573$\\
4 & 0.3665 & 0.0303 & 0.1666 & 71.66 & 0.9855 & $1.2675 \times 10^{-9}$ & $-1.1623$ & $0.2355$ & $0.9276$\\
5 & 0.4067 & 0.0523 & 0.1091 & 66.02 & 1.0432 & $1.6660 \times 10^{-9}$ & $-0.8572$ & $-0.2969$ & $0.9814$\\
6 & 0.4110 & 0.0437 & 0.2645 & 61.07 & 0.9262 & $2.6571 \times 10^{-9}$ & $-1.1786$ & $-0.2929$ & $1.1469$\\
7 & 0.4257 & 0.0496 & 0.0825 & 79.35 & 0.9345 & $2.7991 \times 10^{-9}$ & $-0.9133$ & $0.4121$ & $1.5170$\\
\hline
\end{tabular}
\end{table*}

\subsubsection{Validation Simulations}

In addition to the calibration simulations, we generate 20 independent simulations with random initial phases. These simulations have the same specifications as the primary set ($1\, h^{-1}\, \mathrm{Gpc}$ box size and $1024^3$ particles) but do not use paired phases. They serve as a validation set to test the predictive power of our HMF model. These simulations are not included in the calibration process to ensure an unbiased assessment of the model's accuracy.

\subsection{\label{sec:halofinder}Halo Finder}

For our analysis, we employ the \textsc{Rockstar}\footnote{\url{https://bitbucket.org/gfcstanford/rockstar}} halo finder in combination with the Consistent Trees algorithm.\footnote{\url{https://bitbucket.org/pbehroozi/consistent-trees}} \textsc{Rockstar} is a sophisticated phase-space halo finder that identifies halos by utilizing information from both the spatial and velocity distributions of particles \citep{Behroozi:2011ju}.

The \textsc{Rockstar} algorithm initiates by partitioning the simulation volume into Friends-of-Friends (FoF) groups using a large (0.28) linking length compared to what is commonly used by other FoF-based algorithms ($\sim 0.20$). This partitioning facilitates efficient parallelization. Within each FoF group, \textsc{Rockstar} performs an adaptive hierarchical refinement in six-dimensional phase space (three spatial dimensions and three velocity dimensions), creating a nested set of subgroups. This method allows for precise identification of halos and their substructures, even in densely populated regions of the simulation.

To enhance the temporal consistency and accuracy of the halo properties, we apply the \textsc{Consistent} Trees algorithm \citep{Behroozi:2011js} to the \textsc{Rockstar} catalogs. Consistent Trees tracks halos across consecutive simulation snapshots, ensuring that properties such as halo masses and positions evolve smoothly over time. This dynamical tracking is crucial for reducing scatter and improving the reliability of halo catalogs, which is essential for our modeling.

We adopt the spherical overdensity (SO) definition for halos, with masses defined by an average density within a given sphere, which is equal to the virial overdensity $\Delta_{\rm vir}(z)$ computed with respect to the background density, as described in Sect.~\ref{sec:hmf}. We include all particles within the virial radius in the halo mass calculation, regardless of their gravitational binding.

The halo catalogs produced by \textsc{Rockstar}+\textsc{Consistent} Trees are binned logarithmically based on the number of particles per halo. This binning strategy minimizes the effects of mass discretization and ensures a statistically robust sampling across the halo mass spectrum. 

\subsection{\label{sec:cal}Model Calibration}

This section outlines the Bayesian approach used to calibrate the HMF model. Our methodology closely follows that of~\citet{Euclid:2022dbc}, to which we refer the reader for a comprehensive explanation. Here, we overview the relevant key elements.

The calibration of the HMF involves fitting the theoretical model to the halo counts obtained from our suite of simulations described in Sect.~\ref{sec:sims}. We assume that the number of halos in each mass bin follows a Poisson distribution, motivated by the Press--Schechter formalism~\citep{Press:1973iz}, where halo formation is treated as a random process driven by the collapse of matter overdensities. Therefore, the likelihood of observing $N_i^{\rm sim}$ halos in the $i$-th mass bin at redshift $z$ is given by
\begin{equation}
\ln \mathcal{L}(N_i^{\rm sim} | \pmb{\theta}, z) = N_i^{\rm sim} \ln N_i^{\rm th} - N_i^{\rm th} - \ln(N_i^{\rm sim}!)\,,
\label{eq:poisson_likelihood}
\end{equation}
where $N_i^{\rm th} = N_i(\pmb{\theta}, z)$ is the theoretical prediction obtained by integrating the HMF model over the mass bin and multiplying by the simulation volume, and $\pmb{\theta}$ represents the set of HMF model parameters.

However, in practice, systematic effects and numerical uncertainties can introduce additional scatter, especially in mass bins with many halos. To account for this, we adopt a composite likelihood approach, where we approximate the Poisson distribution with a Gaussian distribution for bins with a sufficiently large number of halos~\citep{Euclid:2022dbc}. The composite log-likelihood is then defined as:
\begin{equation}
\ln \mathcal{L}(N_i^{\rm sim} | \pmb{\theta}, z) =
\begin{cases}
\begin{aligned}
&N_i^{\rm sim} \ln N_i^{\rm th} - N_i^{\rm th} \\
&- \ln(N_i^{\rm sim}!)
\end{aligned} & \text{if } N_i^{\rm sim} \leq 25\,, \\[5ex]
\begin{aligned}
&- \dfrac{1}{2} \left( \dfrac{N_i^{\rm sim} - N_i^{\rm th}}{\sigma_i} \right)^2 \\
&- \dfrac{1}{2} \ln(2\pi \sigma_i^2)
\end{aligned} & \text{if } N_i^{\rm sim} > 25\,,
\end{cases}
\label{eq:composite_likelihood}
\end{equation}
where $\sigma_i$ is the standard deviation, given by:
\begin{equation}
\sigma_i^2 = N_i^{\rm th}\,\left(1 + N_i^{\rm th}\sigma_{\rm sys}^2\right)\,.
\label{eq:likelihood_sigma}
\end{equation}
Here, $\sigma_{\rm sys}$ represents an additional variance component accounting for systematic uncertainties. Differently than~\citet{Euclid:2022dbc}, we leave $\sigma_{\rm sys}$ free to vary in our calibration.

The total log-likelihood used for the calibration is computed by summing over all mass bins, redshifts, and simulation outputs:
\begin{equation}
\ln \mathcal{L}_{\rm total}(\pmb{\theta}) = \sum_{s} \sum_{z} \sum_{i} \ln \mathcal{L}(N_{i,z,s}^{\rm sim} | \pmb{\theta}, z)\,,
\label{eq:total_likelihood}
\end{equation}
where the index $s$ runs over all simulations. We assume that the mass bins, redshifts, and simulations are statistically independent. Although outputs from the same simulation at different redshifts are not strictly independent due to the temporal correlation of structures, we mitigate this effect by selecting snapshots separated by time intervals larger than the typical dynamical time of galaxy clusters (approximately $1.7$ Gyr), following the approach of~\citet{Bocquet:2015pva}.

Assessing robustly the impact of correlation across redshifts would, in principle, require a large number of simulations with the same cosmology but different initial conditions, which goes beyond the scope of this paper. Instead, we performed a post hoc test by re-running the calibration after excluding every other selected redshift. Comparing the results of both calibrations revealed no statistically significant tension, ensuring that our final parameter constraints are robust concerning the chosen redshift binning scheme.

The total log-likelihood is sampled using \textsc{emcee}~\citep{Foreman-Mackey:2012any}.\footnote{\url{https://emcee.readthedocs.io}}$^{,}$\footnote{\url{https://github.com/dfm/emcee}} A total of 896 walkers are used, each sampling 10,000 points. The walkers are set initially around a global maximum found by the \textsc{Dual Annealing} algorithm~\citep{1988JSP....52..479T,1996PhyA..233..395T,1997PhLA..233..216X,PhysRevE.62.4473} implemented in~\textsc{scipy}~\citep{Virtanen:2019joe}.


\section{\label{sec:results}Results}

In this Section, we present our main results, starting with the modeling of the HMF, including the impact of resolution in our simulations in Sect.~\ref{sec:modeling}. We present the final baseline model in Sect.~\ref{sec:model} and its calibration in Sect.~\ref{sec:fit}. The assessment of the robustness is presented in Sect.~\ref{sec:robustness} and a comparison with other models in Sect.~\ref{sec:literature}.

\subsection{\label{sec:modeling}Modeling}

\subsubsection{Resolution effects}

\begin{figure*}
    \centering
    \includegraphics[width=1.0\textwidth]{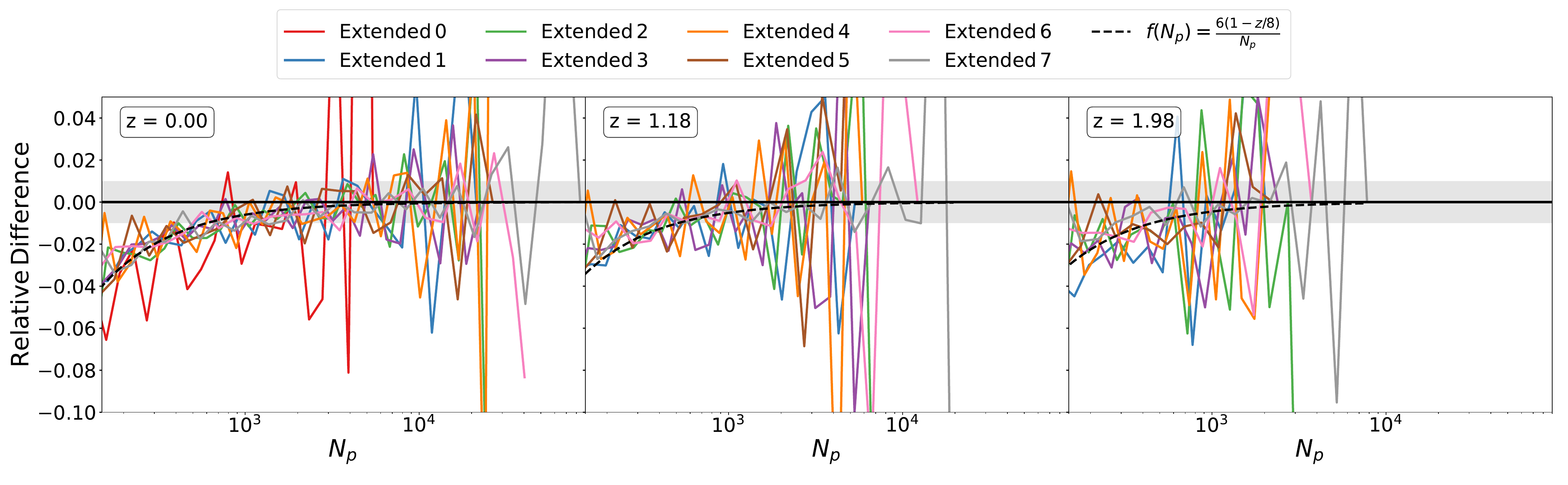}
    \caption{The relative difference of the HMF extracted from the primary set to their high-resolution counterpart. We labeled the cosmologies from the extended set according to their value of $S_8$. The results are shown as a function of the number of particles $N_p$ in the primary simulation.}
    \label{fig:resolution}
\end{figure*}

It is important to assess the convergence of our simulations before modeling the impact of dynamical dark energy on the HMF. In Fig.~\ref{fig:resolution}, we present the relative difference between the HMF extracted from the primary set and their high-resolution counterpart. We present the results as a function of the number of particles $N_p$ in the primary simulation. We note that convergence at the 1 percent level is only achieved at all redshifts for objects with more than a thousand particles in our primary resolution. 

We empirically calibrate the factor by which the HMF needs to be rescaled to account for the effect of resolution, i.e., for the finite number of particles with which a halo is resolved:
\begin{equation}
    f\left(N_p\right) = \frac{6\,(1-z/8)}{N_p}\,,
    \label{eq:res}
\end{equation}
where $z$ is the redshift of a simulation snapshot. This correction factor is shown with the dashed curves in Fig.~\ref{fig:resolution}. We verified that this correction factor is universal, i.e., it depends only on the number of halo particles but not on the simulated cosmology, which is within the range considered here. We also warn that this correction factor is calibrated for the range of redshifts relevant to our analysis. For the rest of the paper, when using the outcomes from the simulations with lower resolution, we will tacitly correct them using Eq.~\eqref{eq:res}. We will also impose a minimum halo mass cut corresponding to 200 particles.

\subsubsection{Baseline model}

In \citet{Euclid:2022dbc}, we demonstrated that the multiplicity function presented in Eq.~\eqref{eq:mult} is sufficiently flexible to accurately fit the results of several scale-free simulations. Consequently, their task of modeling the non-universality of the HMF was reduced to modeling the cosmological dependence of the free parameters in Eq.~\eqref{eq:mult}. We aim to follow a similar approach by starting with a baseline model and then modeling its cosmological dependence. However, the DUCA simulation set offers a significantly larger dynamical range to be fitted, and we found that the same baseline model was not flexible enough to capture the simulation data accurately.

To address this limitation, we assessed the flexibility of different fitting functions by comparing each of them to a single catalog at redshift $z=0$ from one pair of high-resolution simulations. Although the final model will not be calibrated using the high-resolution simulations, this test served as a control to evaluate the overall flexibility of the fitting functions, as the high-resolution simulations provide a larger dynamical range. Empirically, we found that the following fitting function provides accurate results
\begin{equation}
    \nu f(\nu) = 2 A \left\{ \nu_*^r + \left( \dfrac{\nu_p}{\nu_*} \right)^{2p} \right\} \sqrt{ \dfrac{\nu_*^2}{2\pi} } \exp\left( -\dfrac{\nu_*^2}{2} \right) \nu_* ^{q - 1}\,,
    \label{eq:new_mult}
\end{equation}
where $\nu_*=\sqrt{a}\nu$. In the above expression, the expression of the HMF normalization $A$ is given by
\begin{equation}   
\begin{split}
    \left(\frac{A}{\Xi}\right)^{-1} =& -2^{p + \frac{r}{2}} q \, \Gamma\left( \frac{q}{2} + \frac{r}{2} + 1 \right) + 2^{p + \frac{r}{2} + 1} \, p \, \Gamma\left( \frac{q}{2} + \frac{r}{2} + 1 \right) \\
    &- q \, \nu_p^{2p} \, \Gamma\left( -p + \frac{q}{2} + 1 \right) - r \, \nu_p^{2p} \, \Gamma\left( -p + \frac{q}{2} + 1 \right)\,,
\end{split}
\end{equation}
with
\begin{equation}
    \Xi = \frac{\sqrt{\pi} \, (2p - q)(q + r)}{2^{-p + \frac{q}{2} + \frac{1}{2}}}\,.
\end{equation}

Equation~\eqref{eq:new_mult} reduces to Eq.~\eqref{eq:mult} in the limit $r \to 0$ and $\nu_p \to 1$. The rationale behind these modifications is to increase the model's flexibility by introducing a double power-law behavior controlled by the parameter $r$, and incorporating a different pivot scale $\nu_p$.

\subsubsection{Cosmological dependency of $\{a, p, q, r, v_p\}$}

\begin{figure*}
    \centering
    \includegraphics[width=1.0\textwidth]{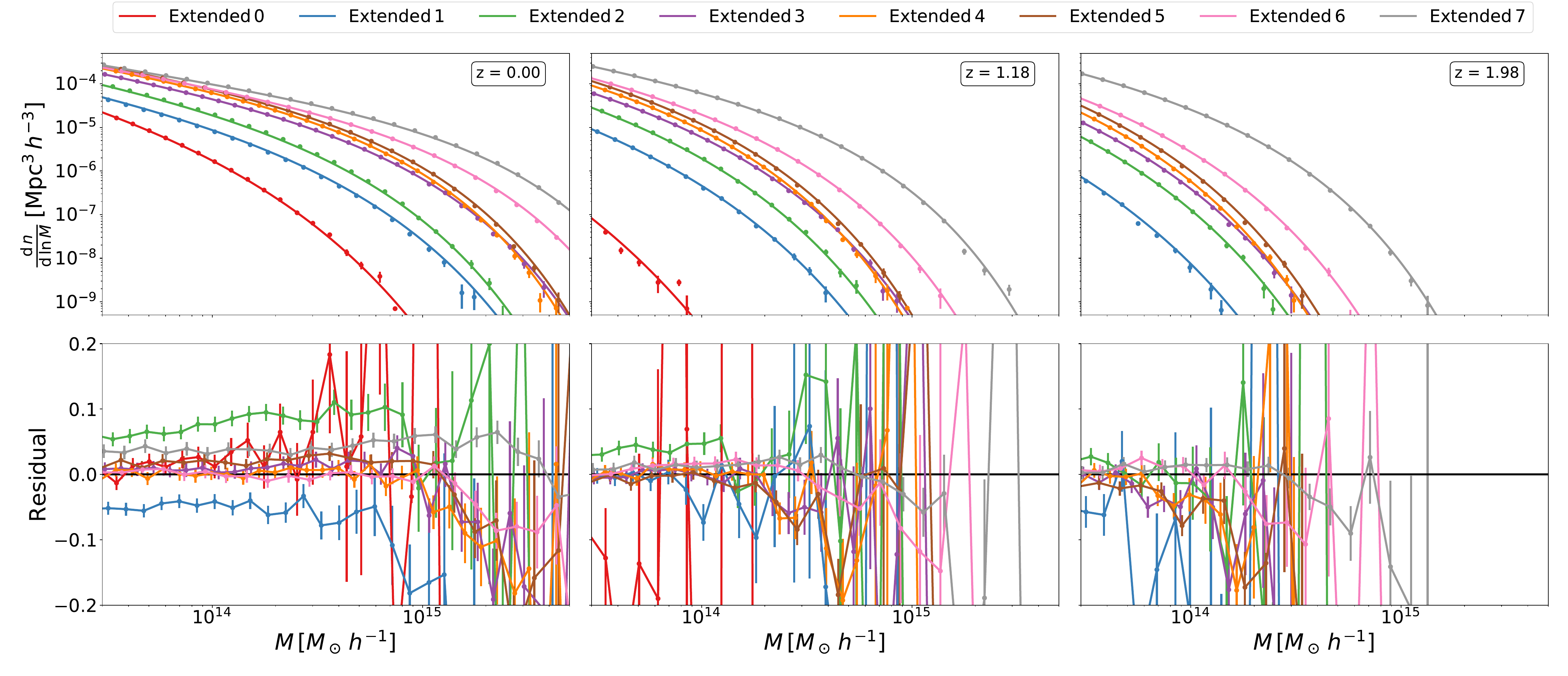}
    \caption{\emph{Top:} The comparison between the HMF predicted by the model of~\citet{Euclid:2022dbc} (curves) and the HMF measured for the extended large simulations (points). Error bars associated with each simulation point correspond to the Poisson uncertainty. \emph{Bottom:} relative difference between the model and the simulations. Different panels indicate different redshifts, from left to right, $z=0.00, 1.18$, and $1.98$.}
    \label{fig:comp}
\end{figure*}

In~\citet{Euclid:2022dbc}, we have shown how the cosmological dependency of the HMF parameters $\{a, p, q\}$ can be modeled as a function of the parameters determining the Friedman background and the shape of the power spectrum. It is natural to impose that the extension allowed by the parameter $r$ behaves similarly to the parameter $p$ (see Eq. \ref{eq:q_param}), as they operate over a similar range in $\nu$. Therefore, we assume that
\begin{equation}
r = r_1 + r_2\,\left(\dfrac{\de \ln \sigma}{\de\ln R} + 0.5\right)\,,
\label{eq:r}
\end{equation}
while we assume $\nu_p$ to be a constant not to introduce unwanted degeneracies with the parameter $a$.

To visually inspect the impact of dynamical dark energy in the HMF, we present in Fig.~\ref{fig:comp}, the comparison between the HMF predictions by the model of~\citet{Euclid:2022dbc} and the observed HMF on the extended large simulations for three different redshifts $z\in\{0.00, 1.18, 1.98\}$. In the top panel, we observe that the model of~\cite {Euclid:2022dbc} qualitatively agrees with the simulations. The differences reported in the bottom panel amount to only few percent even though the simulations analyzed here involve a wider dynamic range and a different range of cosmological parameters with respect to those on which the HMF calibration in~\citet{Euclid:2022dbc} was based. Interestingly, we note that the differences between the model and the simulations are higher at lower redshift, when the background evolution is dominated by dark energy.

From Fig.~\ref{fig:comp}, we conclude that adding a further degree of freedom that couples to the dark energy evolution is needed for the model to absorb the effect of extending the range of validity of our HMF calibration. Clearly, this degree of freedom should only be activated for dynamical dark energy scenarios to build upon the original model. In order to introduce a physically motivated parametrization to account for such a degree of freedom, we consider the quantity
\begin{equation}
    \tilde{q}_{\rm dec}(z, \alpha, \beta) = \frac{3\,\alpha}{2}\,\left[w_{\rm DE}(z)+1\right]\,\Omega_{\rm DE}(z)^\beta\,,
    \label{eq:dof_de}
\end{equation}
where $\alpha$ and $\beta$ are free parameters. For $\tilde{q}_{\rm dec}(z, 1, 1)\equiv\tilde{q}_{\rm dec}(z)$, Eq.~\eqref{eq:dof_de} corresponds to the difference in the deceleration parameter $q_{\rm dec}$ between a cosmology with dynamical dark energy and the corresponding $\Lambda$CDM cosmology with identical present-day values for the density parameters contributed by matter and dark energy.

\begin{figure}
    \centering
    \includegraphics[width=\columnwidth]{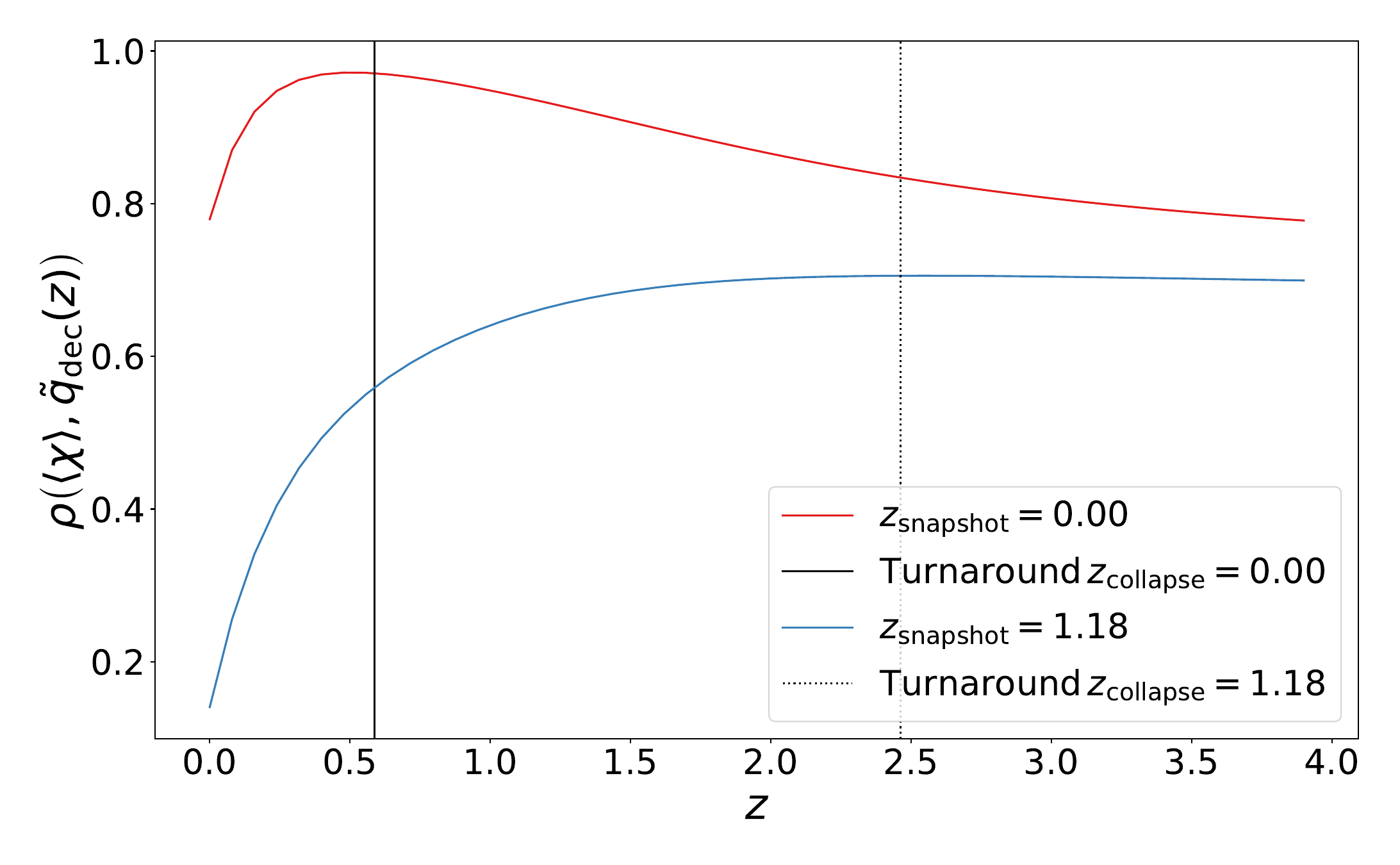}
    \caption{The correlation between the mean residual $\chi$ presented in the bottom panel of Fig.~\ref{fig:comp} with $\tilde{q}_{\rm dec}(z)$ for two snapshots at redshifts $z\in\{0.00, 1.18\}$. We also present the spherical collapse prediction for the turnaround at the Einstein-de Sitter universe for each redshift in solid ($z=0.00$) and dotted ($z=1.18$) black lines.}
    \label{fig:correlation}
\end{figure}

To assess precisely how one should implement the impact of dynamical dark energy, we quantify the correlation between the residuals of~\citet{Euclid:2022dbc} HMF model at a given target redshift $z_{\rm snapshot}$ and the values of the parameter $\tilde{q}_{\rm dec}(z)$ at different past redshifts. Specifically, for each simulation and snapshot, we first compute a single residual, $\chi$, that reflects the average offset (or “mismatch”) between the measured halo abundance and our reference model prediction over the relevant mass bins. Mathematically, if $x_{\rm obs}$ denotes the measured quantity (the ratio of measured to predicted halo counts) and $x_{\rm theo}$ is its theoretical counterpart, then the residual is defined as
\begin{equation}
    \chi = \frac{x_{\rm obs} - x_{\rm theo}}{\sigma_{x,\, \rm error}}\,,
\end{equation}
where $\sigma_{x,\, \rm error}$ accounts for both statistical and systematic uncertainties. We then compare these residuals against $\tilde{q}_{\rm dec}(z)$ at various earlier redshifts $z < z_{\rm snapshot}$. This allows us to identify the epoch at which the effect of dynamical dark energy, encoded by $\tilde{q}_{\rm dec}(z)$, most strongly influences the structures that eventually collapse by $z_{\rm snapshot}$.

To make this comparison quantitative, we compute the correlation coefficient
\begin{equation}
    \rho(\chi, \tilde{q}_{\rm dec}) = \frac{\langle (\chi - \langle \chi \rangle)(\tilde{q}_{\rm dec} - \langle \tilde{q}_{\rm dec} \rangle) \rangle}{\sigma_\chi \,\sigma_{\tilde{q}_{\rm dec}}}\,,
\end{equation}
where $\langle \chi \rangle$ and $\langle \tilde{q}_{\rm dec} \rangle$ are the mean values (across our suite of cosmologies) of the residuals and the deceleration parameter, respectively, and $\sigma_\chi$ and $\sigma_{\tilde{q}_{\rm dec}}$ are their standard deviations. 

Figure~\ref{fig:correlation} illustrates these correlation coefficients for two representative snapshots at $z=0.00$ and $z=1.18$. The vertical lines mark the spherical-collapse turnaround redshift for each snapshot, computed under an Einstein--de Sitter (EdS) assumption. We adopt this EdS approximation to avoid a full numerical solution of the spherical collapse for every cosmology, thereby reducing computational cost while maintaining sufficient accuracy (better than 10\% in the cosmological parameter space we explore). The location of the vertical lines highlights the epoch at which the background expansion balances the collapse, and we see that $\tilde{q}_{\rm dec}(z)$ at this turnaround epoch exhibits the strongest correlation with the residuals of our reference HMF model. This finding supports our choice to incorporate an explicit dependence on the deceleration parameter at turnaround in the final HMF fitting function.

To interpret the correlation between the residual of the reference model predictions and the simulations with $\tilde{q}_{\rm dec}(z)$ at the turnaround redshift, we remind that $\tilde{q}_{\rm dec}(z)$ corresponds to the difference in the background expansion with respect to the static $\Lambda$-dominated expansion. For $w_{\rm DE}<-1$ ($w_{\rm DE}>-1$), the background expansion evolves faster (slower) than the standard model case. As a consequence, structure formation is slowed (accelerated) as gravitational instability has to counterbalance a stronger (weaker) dark energy repulsion. The turnaround is the moment when these two contributions, cosmic expansion and gravitational instability, are in exact balance. In this sense, it is not surprising that the value of $\tilde{q}_{\rm dec}$ at turnaround carries information on the effect of dark energy.

\subsection{\label{sec:model}HMF model}

Equation~\eqref{eq:new_mult} gives our final model for the multiplicity function. For $\Lambda$CDM, the parameters $x_{\Lambda, i}\in\{a, p, q\}$ evolve as described in Eqs.\eqref{eq:a_param}--\eqref{eq:qR_param}, $r$ as described in Eq.~\eqref{eq:r}, and $\nu_p$ is assumed to be a free parameter that does not depend on cosmology. 

The dynamical dark energy dependency for $x_i\in\{a, p, q, r\}$ is accounted for as follows:
\begin{eqnarray}
    x_i & = & x_{\Lambda, i} \, \left\{ 1 + \tilde{q}_{\rm dec}(z_{\rm ta}, \alpha_i, \beta_i)\right\}\\
    & = & x_{\Lambda, i} \, \left\{ 1 + \frac{3\,\alpha_i}{2} \left[w_{\rm DE}(z_{\rm ta})+1\right]\,\Omega_{\rm DE}(z_{\rm ta})^{\beta_i} \right\}\,,
\end{eqnarray}
where $x_{\Lambda, i}$ is the corresponding variable value for the $\Lambda$CDM model, while
\begin{equation}
z_{\rm ta}=\frac{1+z}{(1/2)^{2/3}}-1\,
\end{equation}
is the turn-around redshift for a spherical top-hat perturbation that collapses at redshift $z$ in an EdS cosmology. 

\subsection{\label{sec:fit}Calibration}
\begin{table*}[h!]
\centering
\caption{Maximum likelihood estimates and relative error for the multiplicity function the systematic error parameters.}
\label{tab:MLE_parameters}
\resizebox{\textwidth}{!}{%
\begin{tabular}{ccccccccc}
\hline\hline
$a_1$ & $a_2$ & $a_z$ & $\alpha_{a}$ & $p_1$ & $p_2$ & $q_1$ & $q_z$ & $\sigma$ \\
\hline
\multicolumn{9}{c}{Maximum likelihood estimate} \\
0.8501  & 0.234   & -0.0640  & -0.178  & -0.9880  & -0.49  & 0.559   & 0.02701  & 0.00734 \\
\multicolumn{9}{c}{Error} \\
$1.8\times10^{-4}$  & $3.3\times10^{-3}$ & $3.2\times10^{-4}$ & $2.0\times10^{-3}$ & $4.3\times10^{-3}$ & $1.9\times10^{-2}$ & $9.4\times10^{-3}$ & $7.1\times10^{-4}$ & $8.9\times10^{-5}$ \\
\hline\hline
\end{tabular}%
}
\tablefoot{The parameter errors were estimated from the diagonal of the covariance matrix computed from the \textsc{emcee} chains.}
\end{table*}
\begin{figure*}
    \centering
    \includegraphics[width=\textwidth]{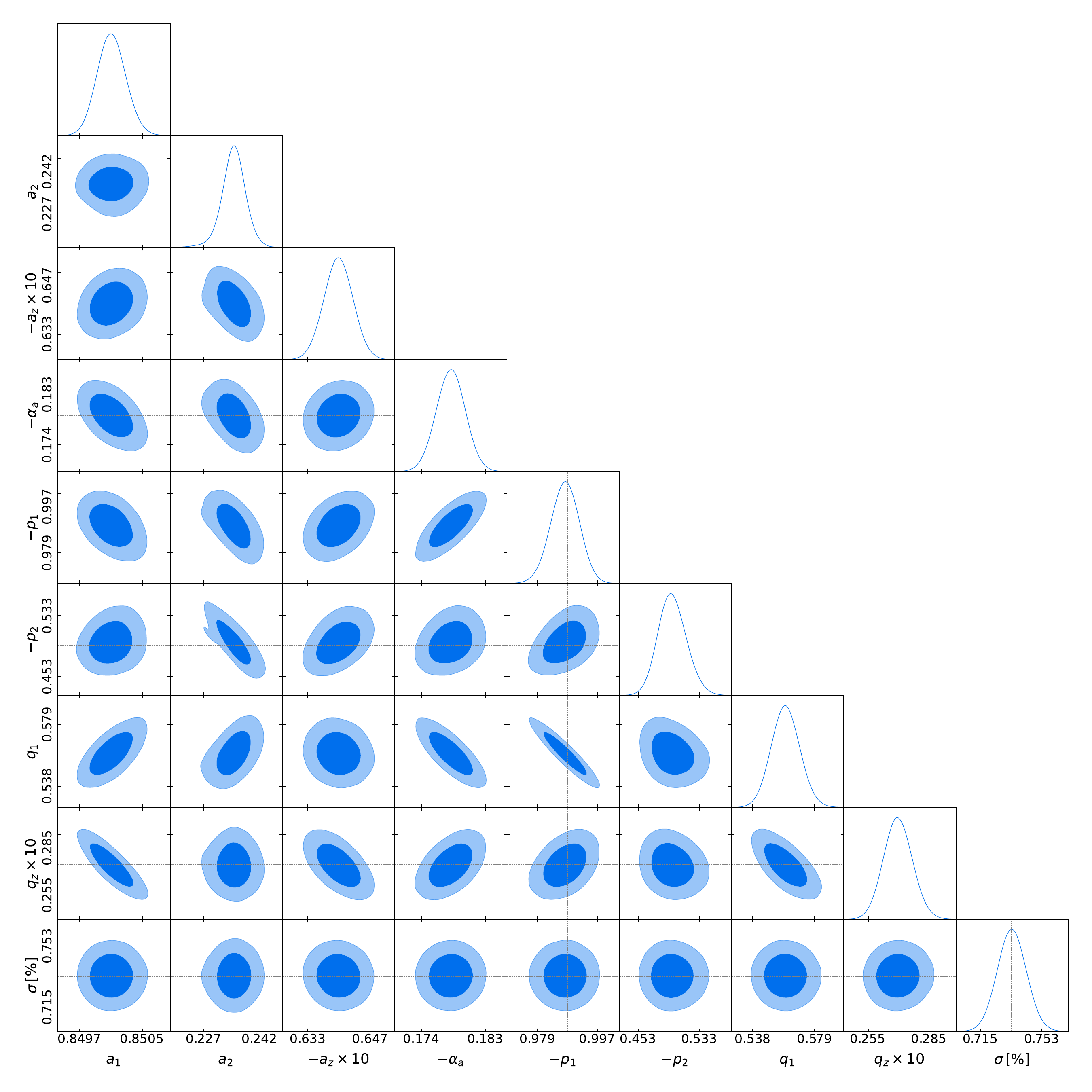}
    \caption{The 68 and 95 percent contour level constraints (dark and light shaded areas, respectively) on the nine free parameters describing the multiplicity function model and the systematic error. Dotted lines mark their values at the maximum likelihood estimate point  (see Table~\ref{tab:MLE_parameters}).}
    \label{fig:chain}
\end{figure*}

When running the first exploratory chains with all parameters free to vary, we detected that only a few parameters were, in fact, statistically independent. To reduce the dimensionality of the model, we have then decided to fix the pair of parameters that presented a correlation coefficient higher than 0.95. They are:
\begin{align}
\alpha_{p} &= 4.7686 \, \alpha_{a} + 0.0051 \\
q_2 &= 1.0020 \, p_2 + 0.7363 \\
r_1 &= -0.9917 \, q_1 + 0.5560 \\
r_2 &= -1.1181 \, p_2 - 0.9405 \\
v_p &= 1.0667 \, q_1 + 1.7577 \,.
\end{align}
Furthermore, all $\beta_i$ parameters were consistent with unity and $\{\alpha_{q},\alpha_{r}\}$ consistent with zero. With all these simplifications, our model has nine parameters free to vary $\{a_1, a_2, a_z, \alpha_a, p_1, p_2, q_1, q_z, \sigma_{\rm sys}\}$. Their maximum-likelihood fit values and relative error are presented in Table~\ref{tab:MLE_parameters}. In Fig.~\ref{fig:chain}, we present the parameters 68 and 95 percent contour level constraints.

Notoriously, our model achieves a sub-percent accuracy in the regime where we are not limited by shot noise as the parameter controlling the systematic error $\sigma$ is smaller than $1\%$. Therefore, our model complies with the calibration requirements for future surveys~\citep[see, for example,][]{Salvati:2020exw,Artis:2021tjj,Euclid:2022dbc}. Lastly, while our model extends the modeling of the HMF to dynamical dark energy, it contains the same number of degrees of freedom for the multiplicity function as its standard model version version~\citep{Euclid:2022dbc}, that is, eight. 

\subsection{\label{sec:robustness}Model Robustness}

\begin{figure}
    \centering
    \includegraphics[width=\columnwidth]{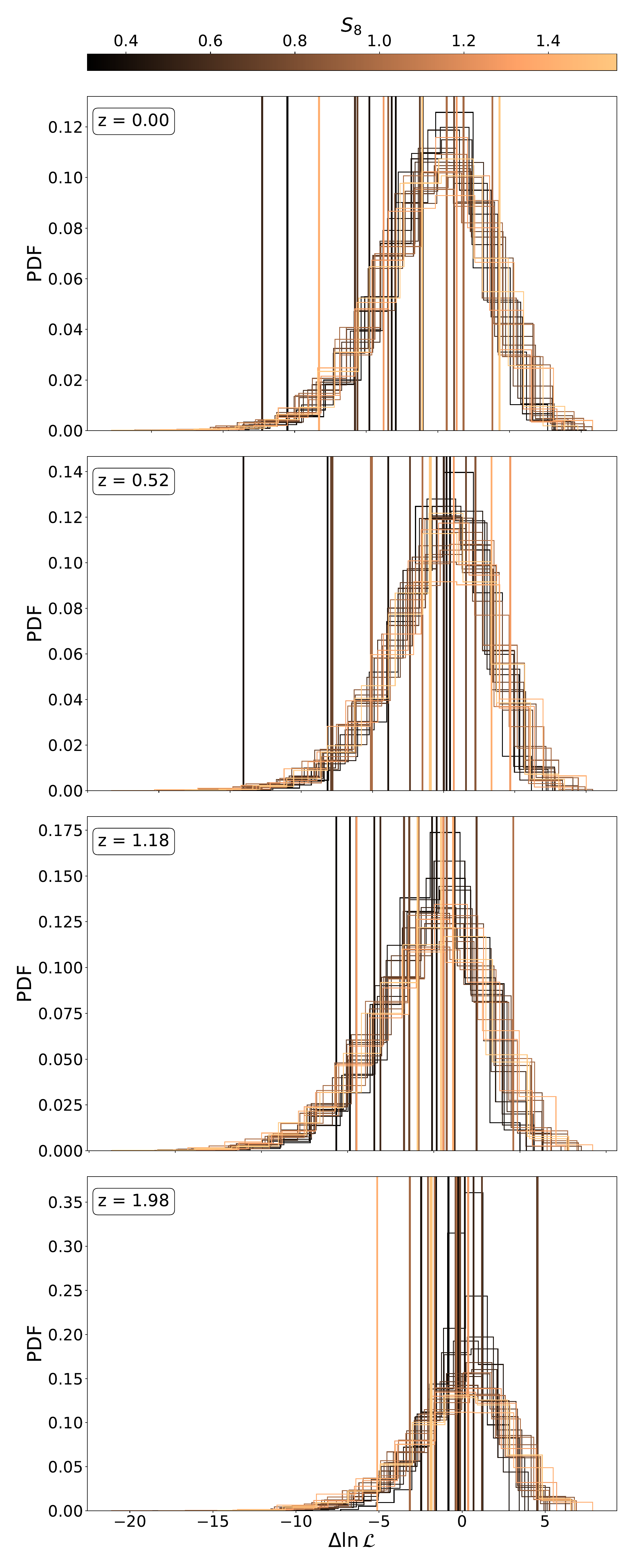}
    \caption{Test $p$-value using the validation set of the DUCA simulations. For each catalog, we have generated $3\times10^5$ synthetic random samples from the likelihood and computed $p$ as the fraction of samples with a likelihood value as extreme as the real catalog. The simulations have been color-coded according to their $S_8$ value. Each vertical line indicates the measured $\Delta \ln \mathcal{L}$ for one of the validation simulations.}
    \label{fig:pvalue}
\end{figure}

\begin{figure*}
    \centering
    \includegraphics[width=\textwidth]{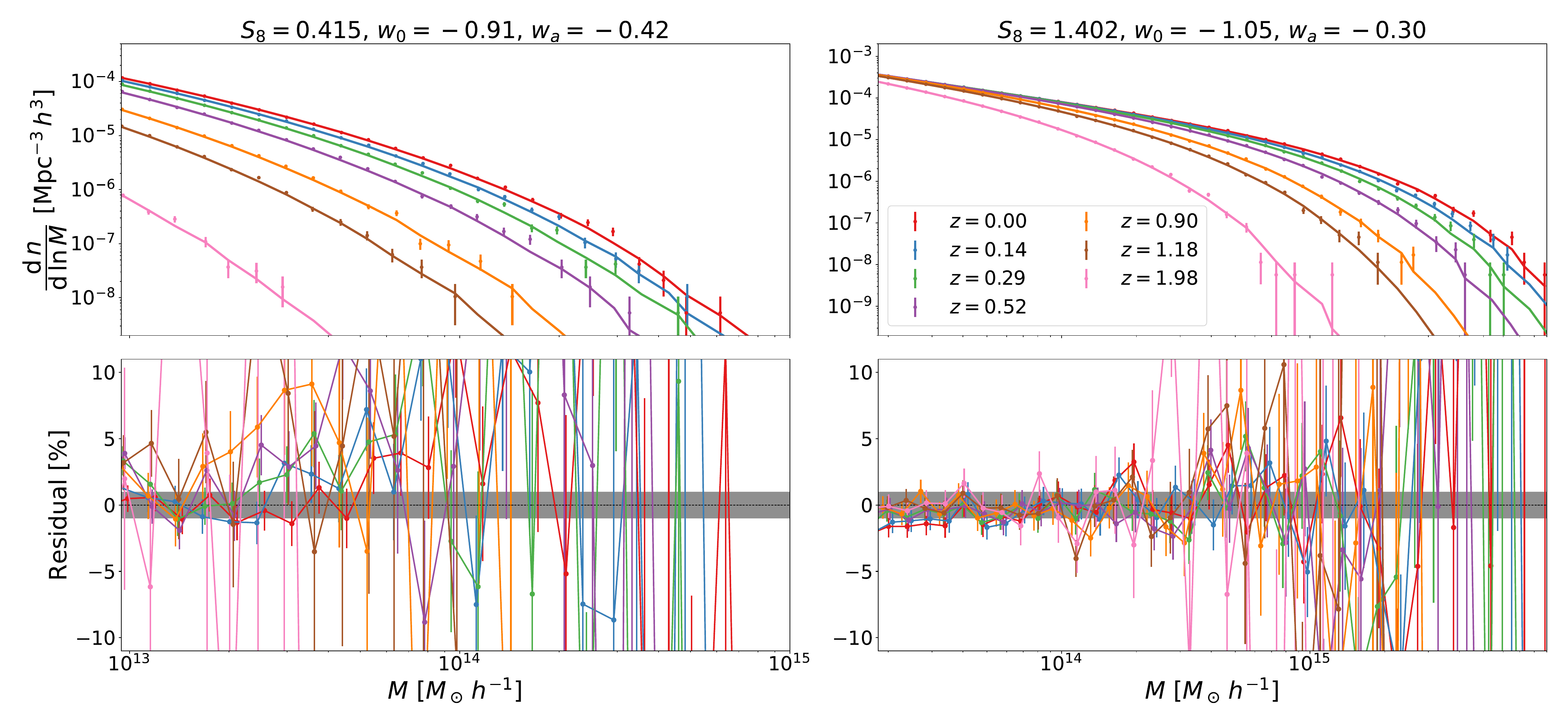}
    \caption{Comparison between the HMF extracted from the two simulations with the \emph{lowest} likelihood values (as shown in Fig.~\ref{fig:pvalue}) and our model predictions. The left panel corresponds to the simulation with the lowest likelihood at redshift $z = 0.52$, and the right panel corresponds to the simulation with the lowest likelihood at $z = 1.18$. Different colors represent different redshifts. The simulation measurements are presented with Poisson error bars.}
    \label{fig:worstcase}
\end{figure*}

To assess our model's robustness, we evaluate a $p$-value test using the validation set of the DUCA simulations. The $p$-value was computed by calculating the likelihood of the HMF extracted from the validation set, assuming our calibrated model. For each catalog, we have generated $3\times10^5$ synthetic random samples from the likelihood and computed $p$ as the fraction of samples with a likelihood value as extreme as the real catalog. The result for all $20$ simulations is shown in Fig.~\ref{fig:pvalue}. To homogenize the distribution of likelihood values, we present the histogram for $\Delta \ln \mathcal{L}$, defined as the likelihood value minus the median of the likelihood value obtained from the synthetic samples. Each vertical line in Fig.~\ref{fig:pvalue} corresponds to one of the validation simulations, and is color-coded according to its $S_8$ value. Lines lying near the histogram’s peak indicate a typical (i.e., likely) outcome under our model, whereas lines falling in the tails signify more extreme results. Thus, the position of a vertical line relative to the histogram reveals how well or poorly that particular simulation’s likelihood agrees with the distribution of synthetic realizations. 

Most catalogs show reasonable values for $\Delta \ln \mathcal{L}$, distributed plausibly around the distribution peak. Two catalogs, one at $z=0.52$ and one at $z=1.18$, distinguish from the sample presenting extreme values for $p$. These two simulations are picked for scrutiny of our model robustness and presented in Fig.~\ref{fig:worstcase}, where we show the model performance for different redshifts. The model on the left corresponds to the simulation with low likelihood at $z=0.52$. It has an extremely low value of $S_8=0.415$, making Poisson noise much more evident in the residual plot in the bottom panel. Still, the model has a percent-level performance for the low-mass objects at low redshifts. More significant deviations appear at higher redshift and mass, but the model is still consistent with the simulation within the error bars for most masses. Specifically for $z=0.52$,  we observe that the mass bins around a few times $10^{14} \msun$ count with significantly more objects than the model. Excluding the two bins around this regime substantially improves the likelihood and the $p$-value. It is important to notice that the original $p$-value for the catalog at this redshift was less than a $3\,\sigma$ fluctuation. Therefore, it is likely that the relatively low performance of the model for this cosmology was just a statistical fluctuation.

The model on the right panel of  Fig.~\ref{fig:worstcase} corresponds to the simulation with low likelihood at $z=1.18$. For this simulation, we observe that the model performance is sub-percent in the regime where we are not limited by Poisson noise, but for the first mass bins at low redshift, the simulation shows a few percent fewer objects than the model. Excluding these mass bins results in a much better likelihood and $p$-value. This case shows that the resolution correction proposed in Eq.~\ref{eq:res} might have a residual cosmological dependency not captured by Eq.~\eqref{eq:res}. This simulation has a high $S_8=1.402$, so a more precise correction should likely consider the higher degree of non-linearity of this specific model. Improving the resolution effect modeling goes beyond the scope and needs of this paper.

From the results of this Section, we conclude that our HMF model incorporating the effect of dynamical dark energy successfully passed an extensive validation, robustly keeping its accuracy and precision level over an extended range for the choice of the relevant cosmological parameters beyond those sampled for the HMF calibration phase.

\subsection{\label{sec:literature}Comparison with other models}

\begin{figure}
    \centering
    \includegraphics[width=\columnwidth]{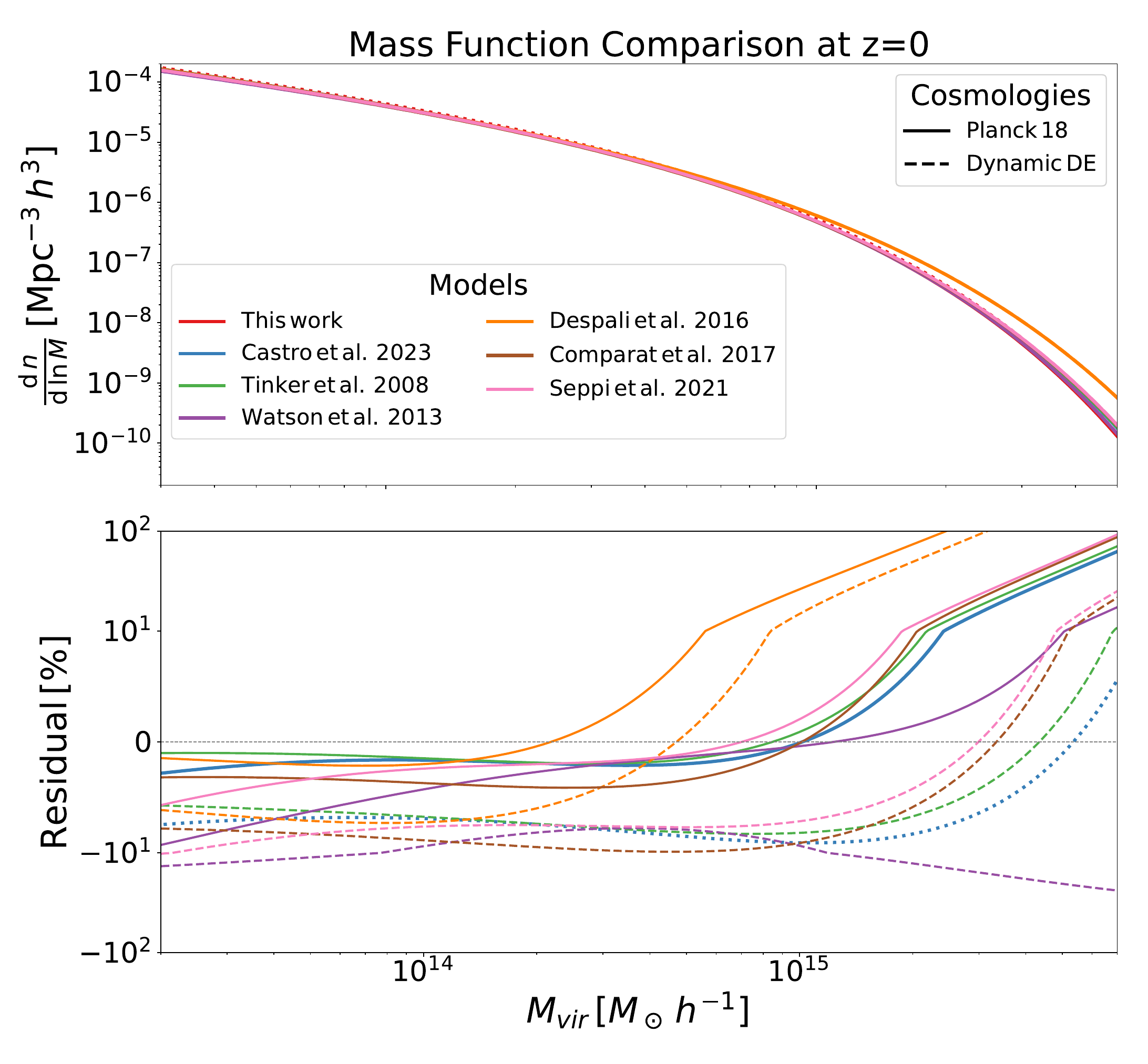}
    \caption{Comparative analysis between our model's prediction for the HMF at $z=0$ and established models in the literature. The comparison is shown for two cosmological scenarios: one assuming cosmological parameters in agreement with Planck 2018 results and one with the same values for all cosmological parameters but $w_0=-0.8$ and $w_a=0.5$. The relative difference between the various models and our HMF calibration is shown in the bottom panel.}
    \label{fig:literature}
\end{figure}

In Fig.~\ref{fig:literature}, we depict a comparative analysis of the HMF at $z=0$, showcasing to what degree our model aligns with or differs from established models in the literature. We show the comparison for two cosmological scenarios, one assuming cosmological parameters in agreement with Planck 2018 results~\citep[labeled Planck 18, see][]{Planck:2018vyg} (solid curves) and one with the exact same values for all cosmological parameters but the dark energy equation of state $w_0$ and $w_a$ that we assume to be $(-0.8, 0.5)$ (dashed curves). In the bottom panel of Fig~\ref{fig:literature}, we present the relative residual between the different models concerning our model. We consider the alternative models for the HMF:~\citet{Euclid:2022dbc}, \citet{Tinker:2008ff}, \citet{Watson:2012mt}, \citet{Despali:2015yla}, \citet{Comparat:2017ejl}, and \citet{Seppi:2020isf} as they adopt the exact same definition for the SO threshold as we do. We use the models as implemented and made publicly available by~\citet{Diemer:2017bwl}.\footnote{\url{https://bdiemer.bitbucket.io/colossus/}}

We observe deviations in the residuals smaller than 10 percent in the regime $10^{13}\lesssim M/M_\odot < 2\times10^{15}$ for all models but~\citet{Despali:2015yla}. The latter shows more significant deviations for $M\gtrsim 4 \times 10^{14} \msun$. Similar results have been reported in~\citet{Euclid:2022dbc}. Notably, the impact of the chosen cosmological model is evident, as shown by the more pronounced spread in residuals, when comparing results under the Dynamic DE scenario. This highlights the lack of sensitivity of the other models to the dark energy equation of state. In contrast, our model proved to maintain its excellent performance in reproducing the predictions from simulations.

\section{\label{sec:conclusions}Conclusions}

In this work, we developed and calibrated a model for the HMF in the context of dynamical dark energy cosmologies, explicitly utilizing the $w_0w_a$CDM framework, i.e. using the CPL phenomenological parameterization of the dark energy equation of state ~\citep{Chevallier:2000qy,Linder:2002et}, as provided by Eq. \ref{eq:w_of_z}. We extended our previous modeling presented in~\citet{Euclid:2022dbc} by including an additional degree of freedom to account for the evolving nature of dark energy. By leveraging the DUCA $N$-body cosmological simulations, which were carefully designed with a wide range of cosmological parameters, we have assessed the impact of dynamical dark energy on the HMF and proposed a new fitting function that effectively incorporates these effects.

We demonstrated that the dynamical dark energy component introduces a significant dependence on cosmology in the HMF parameters (see Fig.~\ref{fig:comp}). This dependency is well captured by an enhanced parametrization of the multiplicity function that introduces an additional dynamical degree of freedom through the value that the deceleration parameter $\tilde{q}_{\rm dec}$, defined in Eq.~\eqref{eq:dof_de} takes at the epoch at which a spherical top-hat perturbation, collapsing at a given time,  earlier reaches its turnaround (see Fig.~\ref{fig:correlation}).

The robustness of our model was assessed using a validation set of $20$ independent simulations spanning a wide range of cosmologies (see Fig.~\ref{fig:pvalue}). The results consistently showed that our model maintained its sub-percent accuracy across this set, thus comparable to the residuals found for the calibration set. Despite minor deviations in a few extreme cosmologies, particularly those with very high or very low $S_8$ values that we presented in Fig.~\ref{fig:worstcase}, the model proved effective, confirming that our new formulation is robust and broadly applicable to various cosmological models.

Furthermore, in Fig.~\ref{fig:literature}, our comparative analysis demonstrated that existing HMF models introduced in the literature cannot accurately reproduce the effects of including dynamical dark energy. Our model, in contrast, consistently maintained its precision, effectively capturing the nuances introduced by the evolving dark energy equation of state. The validation and comparison results suggest that our model is well-suited to enhance the cosmological information carried by galaxy clusters identified in undergoing and upcoming multi-wavelength surveys.


\section{\label{sec:data}Data availability}

In~\citet{Castro_CCToolkit_A_Python_2024}, we implement the new model presented in this paper. The source code can be accessed in~\url{https://github.com/TiagoBsCastro/CCToolkit}.


\begin{acknowledgements}

It is a pleasure to thank Duca dos Anjos. We are grateful to thank Isabella Baccarelli, Fabio Pitari, and Caterina Caravita for their support with the CINECA environment. The DUCA simulations were run on the Leonardo-DCGP supercomputer as part of the Leonardo Early Access Program (LEAP). TC and SB are supported by the Agenzia Spaziale Italiana (ASI) under - Euclid-FASE D  Attivita' scientifica per la missione - Accordo attuativo ASI-INAF n. 2018-23-HH.0, by the PRIN 2022 PNRR project "Space-based cosmology with Euclid: the role of High-Performance Computing" (code no. P202259YAF), by the Italian Research Center on High-Performance Computing Big Data and Quantum Computing (ICSC), a project funded by European Union - NextGenerationEU - and National Recovery and Resilience Plan (NRRP) - Mission 4 Component 2, by the INFN INDARK PD51 grant, and by 
the National Recovery and Resilience Plan (NRRP), Mission 4,
Component 2, Investment 1.1, Call for tender No. 1409 published on
14.9.2022 by the Italian Ministry of University and Research (MUR),
funded by the European Union – NextGenerationEU– Project Title
"Space-based cosmology with Euclid: the role of High-Performance
Computing" – CUP J53D23019100001 - Grant Assignment Decree No. 962
adopted on 30/06/2023 by the Italian Ministry of Ministry of
University and Research (MUR);
We acknowledge the computing centre of CINECA and INAF, under the coordination of the ``Accordo Quadro (MoU) per lo svolgimento di attività congiunta di ricerca Nuove frontiere in Astrofisica: HPC e Data Exploration di nuova generazione'', for the availability of computing resources and support. We acknowledge the use of the HOTCAT computing infrastructure of the Astronomical Observatory of Trieste -- National Institute for Astrophysics (INAF, Italy) \citep[see][]{2020ASPC..527..303B,2020ASPC..527..307T}.
\end{acknowledgements}

\bibliography{mybib}

\end{document}